\documentclass[useAMS,usenatbib,usegraphicx]{mn2e}
\usepackage{amsmath,amssymb}
\usepackage{graphicx,mathptm}
\usepackage{times}
\usepackage{natbib}
\usepackage{epsfig}

\newcommand{\F}{$\mathit{F}$}
\newcommand{\mh}{$M_{\rm Halo}$}

\title[Merger bias and quasar clustering] {On merger bias and the
clustering of quasars}

\author[Bonoli et al.]
{Silvia Bonoli$^1$\thanks{E-mail:bonoli@mpa-garching.mpg.de},
Francesco Shankar$^1$,
Simon D.M. White$^1$,
Volker Springel$^1$,
\newauthor
J. Stuart B. Wyithe$^2$ 
\\
$^1$Max-Planck-Institut f\"ur Astrophysik, Karl-Schwarzschild Strasse 1,
D-85740 Garching, Germany\\
$^2$School of Physics, University of Melbourne, Parkville, Victoria, Australia
\vspace{-0.5cm}
}

\begin{document}



\maketitle

\label{firstpage}

\begin{abstract}
We use the large catalogues of haloes available for the Millennium
Simulation to test whether recently merged haloes exhibit stronger
large-scale clustering than other haloes of the
same mass. This effect could help to understand the very strong
clustering of quasars at high redshift. However, we find no statistically
significant excess bias for recently merged haloes over
the redshift range $2\le z \le 5$, with the most massive haloes showing
an excess of at most $\sim 5\%$.  We also consider galaxies
extracted from a semianalytic model built on the Millennium
Simulation.  At fixed stellar mass, we find  an excess
bias of $\sim 20-30\%$ for recently merged objects,
decreasing with increasing stellar mass. The
fact that recently-merged galaxies are found in systematically
 more massive haloes than other galaxies of the same stellar mass accounts for
about half of this signal, and perhaps more for high-mass galaxies. 
The weak
merger bias of massive systems suggests that objects of merger-driven nature,
such as quasars, do not cluster significantly differently than other objects of
the same characteristic mass.  
We discuss the implications of these
results for the interpretation of clustering data with respect to quasar
duty cycles, visibility times, and evolution in the black hole-host
mass relation.
\end{abstract}


\section{Introduction}\label{sec:intro}

In the last decade, much theoretical work has tried to constrain the
cosmological evolution of supermassive black holes (BHs) by
simultaneously interpreting the statistics of quasars/BHs and their
clustering as a function of redshift and luminosity
\citep[e.g.,][]{kauffmann02, WL05, lidz06, hopkins07b, thacker08,
  bonoli09,ShankarCrocce, SWM}.  In fact, if quasars are hosted by
haloes whose bias is only mass-dependent, clustering measurements can be
used to infer the mass \mh\ of the host dark matter halo, which in turn
provides host number densities, quasar duty cycles 
(here defined as the ratio between quasar
and halo number densities) and scatter in the relation between quasar
luminosity $L$ and halo mass \citep{cole89, haehnelt98, martini01,
  haiman01}.  Measuring a high bias implies high halo masses, low host
number densities, high duty cycles, and vice versa. At fixed duty cycle,
increasing the scatter in the mean $L$-\mh\ relation implies increasing
the contribution of less massive and less biased haloes to the same
luminosity bin, thus lowering the overall bias.

The advent of wide-field surveys like the Sloan Digital Sky Survey
(SDSS) and the 2dF quasi-stellar object (2dFQSO) \citep{york00,croom04}
survey, with the detection of thousands of quasars, has allowed a
detailed investigation of the clustering properties of accreting BHs
from the local universe up to $z\sim5$ \citep[e.g.,][]{porciani04,
  croom05, shen07, myers07, coil07, daAngela08,padmanabhan08, ross09}.
Assuming that haloes hosting quasars are typical in the way they trace the dark
matter density field, these studies concluded that quasars reside at all
times in a relatively narrow range of halo masses,  \mh$\sim
3\times 10^{12} -  10^{13} h^{-1} \rm{M}_{\odot}$.

Interestingly, the very high clustering amplitude of luminous quasars at
$z>3$ measured with the SDSS \citep{shen07} has posed some nagging
theoretical problems for the simultaneous interpretation of the
clustering and the luminosity function at these epochs. The high
clustering appears to imply that the quasars live in very massive
haloes. But the extreme rareness of such haloes is difficult to
reconcile with the observed quasar luminosity function, especially at
$z\sim 4$, unless a high quasar duty cycle and a very low scatter in the
$L$-\mh\ relation are assumed \citep{White08}.  Moreover, matching the
high $z \gtrsim 3-4$ quasar emissivity to the low number density of
hosts constrains the ratio of the radiative efficiency of accretion
$\epsilon$ to the Eddington ratio $\lambda$ to be $\epsilon>
0.7\lambda/(1+0.7\lambda)$, implying $\epsilon > 0.17$ for $\lambda >
0.25$ \citep{ShankarCrocce}, which are rather extreme values.  However,
these conclusions can be relaxed if quasar hosts cluster more strongly
than typical haloes of similar mass \citep{wyithe09}. This would then
imply that quasars live in less massive but more numerous haloes,
allowing for lower duty cycles and less extreme values for~$\epsilon$.

Several analytical and numerical studies have investigated whether haloes of
similar mass have different large-scale clustering properties,
depending, in a non-trivial way, on their growth history, concentration,
spin, or environment
\citep[e.g.,][]{Kolatt99,Lemson99,KH02,Gao05,Wechsler06,Gao07, angulo08}.  In
particular, \citet{wyithe09} suggest that the possible merger-driven
nature of quasars might cause an excess bias, if the large-scale
clustering of recently-merged haloes is higher than 
expected for typical haloes of the same mass (``merger bias'').  This
suggestion was  based on the model by \citet{FK06}, who
analytically calculated that close merging pairs might possess a merger
bias of a factor of $\gtrsim 1.5$.  

Recent work has indeed shown that clustering strength depends not only on halo mass,
but also on other parameters.
\citet{Gao05} found that later forming haloes with mass $M<M*$ 
are less clustered than typical haloes of the same mass (``assembly bias'').
\citet{Wechsler06} extended this study to show that less concentrated
haloes more massive than the non-linear mass scale are instead more
biased than average. \citet{li08} explored various definitions of halo
formation time, and concluded that the dependence of clustering on 
halo history depends strongly on the precise aspect of the history that
is probed: while they confirmed previous results on assembly bias,
they did not find any dependence of clustering on the time of the last
major merger. Other numerical work that specifically looked at 
merger bias found inconclusive results, probably due to the different
ranges of masses, redshifts and scales used and the poor statistics
available \citep{gottlober02, percival03, scannapieco03}.
\citet{Wetzel07} used a large dark matter simulation to study the
clustering of very massive haloes ($M_{\rm Halo} > 5.0\,
\times 10^{13}\, h^{-1}\, \rm{M}_{\odot}$ ), and found that, at redshift $z
\lesssim 1$, merger remnants show an excess bias of $\sim 5-10\%$.

In this work we make use of the large, publicly available catalogues of
the Millennium Simulation \citep{springel05b} to explore for a wider
range of masses the level of excess bias for high redshift
merger remnants.  In Section~\ref{sec:method} we briefly describe the
simulation and how we identify recent mergers both of haloes and
galaxies. In Section~\ref{sec:results} we show results for the
merger bias both of haloes and of galaxies, and in
Section~\ref{sec:qso_clustering} we discuss the implications of
our results for the clustering of quasars. A summary and our conclusions
are presented in Section~\ref{sec:summary}.

\section{Identifying merging haloes and galaxies in the Millennium Simulation}
\label{sec:method}

\subsection{The Millennium Simulation and its galaxy population}

The Millennium Simulation \citep{springel05b} is an N-body simulation
which follows the cosmological evolution of $2160^{3} \simeq 10^{10}$
dark matter particles, each with mass $\sim 8.6 \times 10^{8} h^{-1}
\rm{M}_{\odot}$, in a periodic box of $500 h^{-1} \rm{Mpc}$ on a
side. The cosmological parameters used in the simulation are consistent
with the WMAP1 \& 2dFGRS `concordance' $\Lambda$CDM framework:
$\Omega_{m} = 0.25$, $\Omega_{\Lambda} = 0.75$, $\sigma_{8} =0.9$,
Hubble parameter $h=H_{0}/100 ~ \rm{km} \rm{s}^{-1} \rm{Mpc}^{-1} =0.73$
and primordial spectral index $n=1$ \citep{spergel03}. 
In the present work we focus on the clustering of galaxies and haloes
from $z=2$ to $z=5$, where the time between two
simulation outputs varies from approximately $200 ~ \rm{Myr}$ to $100 ~
\rm{Myr}$.  This time resolution is good enough to capture
merger events reliably and in these time intervals any change in the large-scale
 distribution of merger remnants is negligible.

Detailed merger trees were constructed for the simulation by identifying
haloes and subhaloes with, respectively, a friends-of-friends (FOF)
group-finder and an extended version of the {\small SUBFIND} algorithm
\citep{springel01}: particles are included in the same FOF group if
their mutual separation is less than $0.2$ of the mean particle
separation. The {\small SUBFIND} algorithm then identifies locally
overdense and self-bound particle structures within FOF groups to
isolate bound subhaloes (which are required to contain a minimum of $20$
particles).  For further details on the Millennium Simulation and the
tree building procedure we refer the reader to \citet{springel05b}.

The formation and evolution of galaxies has been followed in a
post-processing simulation which uses the dark matter merger trees as
basic input combined with analytical treatments of the most important
baryonic physics in galaxy formation, including the growth of central BHs 
\citep{croton06a, Bower06,
  delucia07}. This has produced remarkably successful galaxy formation
models which reproduce a large set of observational findings about the
local and high redshift galaxy populations with good accuracy. While not
perfect, this match justifies substantial trust in the basic paradigm of
hierarchical galaxy formation in CDM cosmologies, and motivates detailed
studies of the merger and clustering statistics using the Millennium
Simulation.  Below we describe our definition of major mergers both for
dark matter haloes and galaxies, which lies at the heart of our
investigation of the merger bias phenomenon.

\subsection{Halo mergers}
\label{sec:halomergers}

We note that different definitions of halo formation time have led to
somewhat different quantitative conclusions regarding the effect of
assembly history on the large-scale clustering of haloes
\citep[e.g.,][]{Gao05,Wechsler06, li08}. Here we are interested in
the possible bias caused by recent merger activity, which might induce a
non-trivial relation between the clustering of dark matter haloes and
objects whose formation is triggered by mergers (such as quasars).  We
are therefore not interested in tracking the full mass accretion history
of dark matter haloes, but rather want to focus on the violent major
merger events that are thought to trigger efficient BH accretion and
starburst activity.

In the present work, we consider as {\em major mergers} those events in
which two separate haloes with comparable masses encounter each other
for the first time, that is, when they join the same FOF group. At a
given time $z_{n}$, we define as recently merged haloes those that, at
the previous snapshot $z_{n-1}$, have two or more progenitors belonging to separate
FOF groups whose mass (defined through the number of particles of the
FOF group) was $> 20\% $ that of the descendant (corresponding to a
minimum ratio $m_{\rm sat}:m_{\rm central}=1:4$). This definition of a
major merger is similar to the one of \citet{scannapieco03}, who defined
as merger remnants haloes that, within the time interval of a single
snapshot, accreted at least $20\%$ of their final mass. These authors,
however, also considered haloes that experienced considerable smooth
mass accretion, whereas we strictly require the merger remnant to be the
product of the encounter of two sufficiently massive FOF progenitors. We
note that \citet{Wetzel07} used different definitions of a major merger,
finding their results to be independent of the exact definition, as long
as a substantial amount of mass is accreted in a relatively short
time. We therefore also expect our results to be robust with
respect to the specific value of the mass ratio threshold adopted here.

In our definition of major mergers we also include encounters of groups
that at some later time might split again. This can occasionally happen
since the FOF algorithm sometimes links two haloes that are just passing
close to each other but that in the future will (at least temporarily)
separate again.  To check whether this might impact our overall results,
we also used the merger trees extracted from the Millennium Simulation
by \citet{genel08}, who carefully excluded all mergers of FOF groups
containing subhaloes that at a future time will belong to two different
FOF groups.  Moreover, \citet{genel08} define as halo mass the sum of
just the gravitationally bound particles.  We checked, however, that our
results do not change when switching to the \citet{genel08} halo trees.
The differences from our reference catalogues affect the halo population
only at very low redshifts and at low halo masses, much below the ranges
of interest here.

We have also checked that our definition of halo mass based on the
number of linked particles instead of a spherical overdensity mass
estimate does not affect our result. As an alternative to the FOF group
masses, we used as group masses the mass within the radius that encloses
a mean overdensity of $200$ times the critical density, or the mass
within the radius where the overdensity is that expected for
virialization in the generalized top-hat collapse model for our
cosmology. However, we found that this did not lead to any significant
differences in the large-scale clustering properties of haloes as a
function of mass.


\subsection{Galaxy mergers}
\label{sec:galmergers}


In the Millennium Simulation, the orbits of dark matter subhaloes are followed
until tidal truncation and stripping due to encounters with larger
objects cause them to fall below the simulation resolution limit ($20$
particles, equivalent to a mass of $\sim 1.7\times 10^{10}\, h^{-1}\,
\rm{M}_{\odot}$). Galaxies follow the orbits of their host subhalo until
this point, and then their remaining survival time as satellite galaxies
is estimated using their current orbit and the dynamical friction
formula of \citet{binney87}, calibrated as in \citet{delucia07}. At the
end of this interval, a satellite galaxy is
assumed to merge with the central galaxy of the host dark matter
halo, which can either be a subhalo or, more frequently, the main halo
of the associated FOF group \citep{angulo08b}.

In the event of a minor galaxy merger, the cold gas of the satellite
galaxy is transferred to the disc component of the central galaxy
together with the stars produced in a starburst (as described below);
moreover, the bulge of the central galaxy grows by incorporating all the
stars of the satellite.  If instead a major galaxy merger has occurred,
the discs of both progenitors are destroyed and all stars in the merger
remnant are gathered into the spheroidal bulge component.  In the galaxy
formation model studied here, the starbursts induced by galaxy mergers
are described using the ``collisional starburst'' prescription
introduced by \citet{somerville01}: the fraction $e_{\rm burst}$ of cold
gas which is converted into stars in the merger remnant is given by:
$e_{\rm burst}=\beta_{\rm burst}(m_{\rm sat}/m_{\rm
  central})^{\alpha_{\rm burst}}$, where $\alpha_{\rm burst}=0.7$ and
$\beta_{\rm burst}=0.56$, chosen to provide a good fit to the numerical
results of \citet{cox04}.

We define as major merger remnants those galaxies that have, in the
immediately preceding simulation output, two progenitors with stellar
masses larger than $20\%$ of the stellar component of the descendant (as
for the FOF haloes, this imposes a minimum mass ratio $m_{\rm
  sat}:m_{\rm central}=1:4$). Note that this definition is close, but
not identical, to the distinction between minor/major mergers adopted in
the underlying galaxy formation model.

\begin{figure*}
\begin{center}
	 \includegraphics[width=0.8\textwidth]{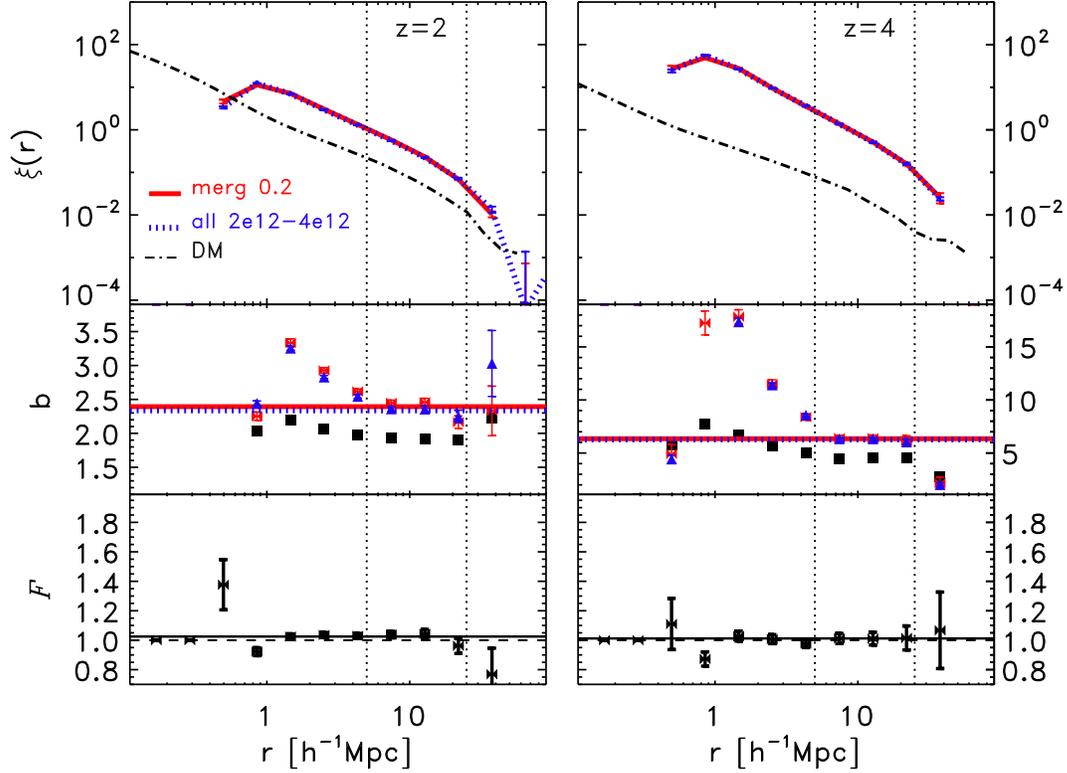}
	\caption{\emph{Upper panels}: Examples of the two-point
          cross-correlation function at $z=2$ and $z=4$ (left and
          right, respectively) between the reference sample and the
          haloes with mass in the range $2.0 < M_{\rm Halo} < 4.0\,
          \times 10^{12}\, h^{-1}\, \rm{M}_{\odot}$ (blue-dotted
          lines), and between the reference sample and the sub-sample
          of recently merged haloes (red lines). The auto-correlation
          of the underlying dark matter is shown as dot-dashed
          line.  \emph{Middle panels}: Bias as a function of scale for
          all the haloes in the mass bin (blue triangles), for the
          corresponding merger remnants (red bow-ties) and for the
          reference sample (black squares).  The horizontal lines
          indicate the fit to the points, over the range indicated by the
	  vertical dotted lines..  \emph{Lower panels}: Excess
          bias $\mathit{F}$ for the merger remnants relative to the whole
          galaxy population as a function of scale. The horizontal
          dashed line indicates 
          $\mathit{F}=1$.  We refer the reader to the text for
          details of the calculation of errors and the fitting
          procedure.}
	\label{fig:correlation_haloes}
\end{center}
\end{figure*}

\section{RESULTS}
\label{sec:results}

\subsection{Clustering analysis and the excess bias $\mathit{F}$}

We use the standard definition of the {\em two-point spatial correlation
  function} as the excess probability for finding a pair of objects at a
distance {\em r}, each in the volume elements ${\rm d}V_{1}$ and ${\rm
  d}V_{2}$ \citep[e.g.,][]{peacock99}:
\begin{equation}\label{eqn:two-point}
{\rm d}P=n^{2} \left [ 1+ \xi(r) \right ] {\rm d}V_{1} {\rm d}V_{2},
\end{equation}
where $n$ is the average number density of the set of objects under consideration.

The {\em bias} between two classes of objects (e.g., haloes and dark
matter) is defined as the square-root of the ratio of the corresponding
two-point correlation functions:
\begin{equation} \label{eqn:bias}
b_{ H,\, \rm{DM}} (r) \equiv \sqrt{\frac {\xi_{ H} (r)} {\xi_{\rm DM} (r)}}.
\end{equation}

Since the number density of merging objects at a given snapshot is too
low for a statistically significant auto-correlation study (see section
\ref{sec:qso_clustering}), we adopt a cross-correlation analysis
instead. In this case the bias is given by
\begin{equation}\label{eqn:crossbias}
b_{ H,\, \rm{DM}} (r) \equiv \frac{1}{b_{ R, \, \rm{DM}}(r)} \frac{\xi_{H,\,
R}(r)}{\xi_{\rm DM}(r)} ,
\end{equation}
where $b_{ R, \,\rm{DM}} (r)$ is the bias (relative to the dark matter)
of the population we are using as reference in our cross-correlation
analysis, and $\xi_{ H,\,R}(r)$ is the cross-correlation function
between the haloes and the reference population.  By definition, the
bias is a function of scale. However, the scale dependence becomes weak
or even vanishes at large scales. Since we are here interested in the
behavior of the merger bias at very large scales, we estimate the bias
on these scales by finding the best constant value over the range $5 <
r < 25 ~ h^{-1} \, \rm{Mpc}$. This adds robustness to our results by reducing
noise from counting statistics.

We can define the merger bias as the excess in the clustering of merger
remnants at large scales with respect to the global population of
objects selected with similar properties:
\begin{equation}
\mathit{F}(r) = \xi_{M, \, R}(r)/\xi_{H, \, R}(r),
\end{equation}
where $\xi_{M, \, R}$ is the cross-correlation between merger remnants
and the reference sample, and $\xi_{H, \, R}$ is the cross-correlation
between the global population and the reference sample.

\begin{figure}
\begin{center}
	 \includegraphics[width=0.48\textwidth]{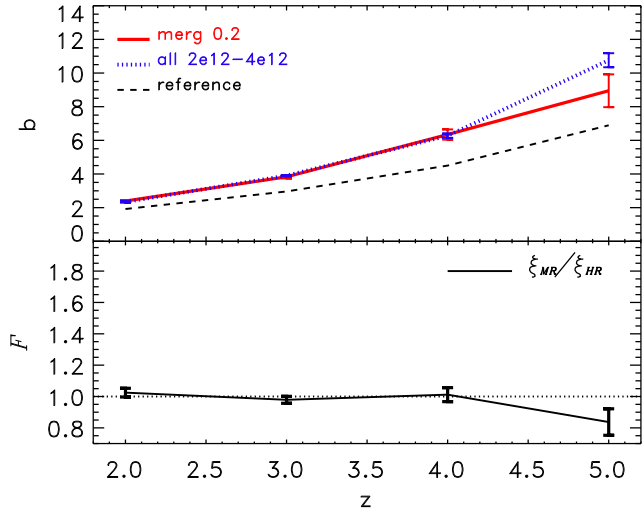}
	\caption{Bias and $\mathit{F}$ parameter as a function of
          redshift, from the best fit obtained for the halo samples shown
          in Figure~\ref{fig:correlation_haloes}.}
	\label{fig:bias_f}
\end{center}
\end{figure}

\begin{figure}
\begin{center}
	\includegraphics[width=0.49\textwidth]{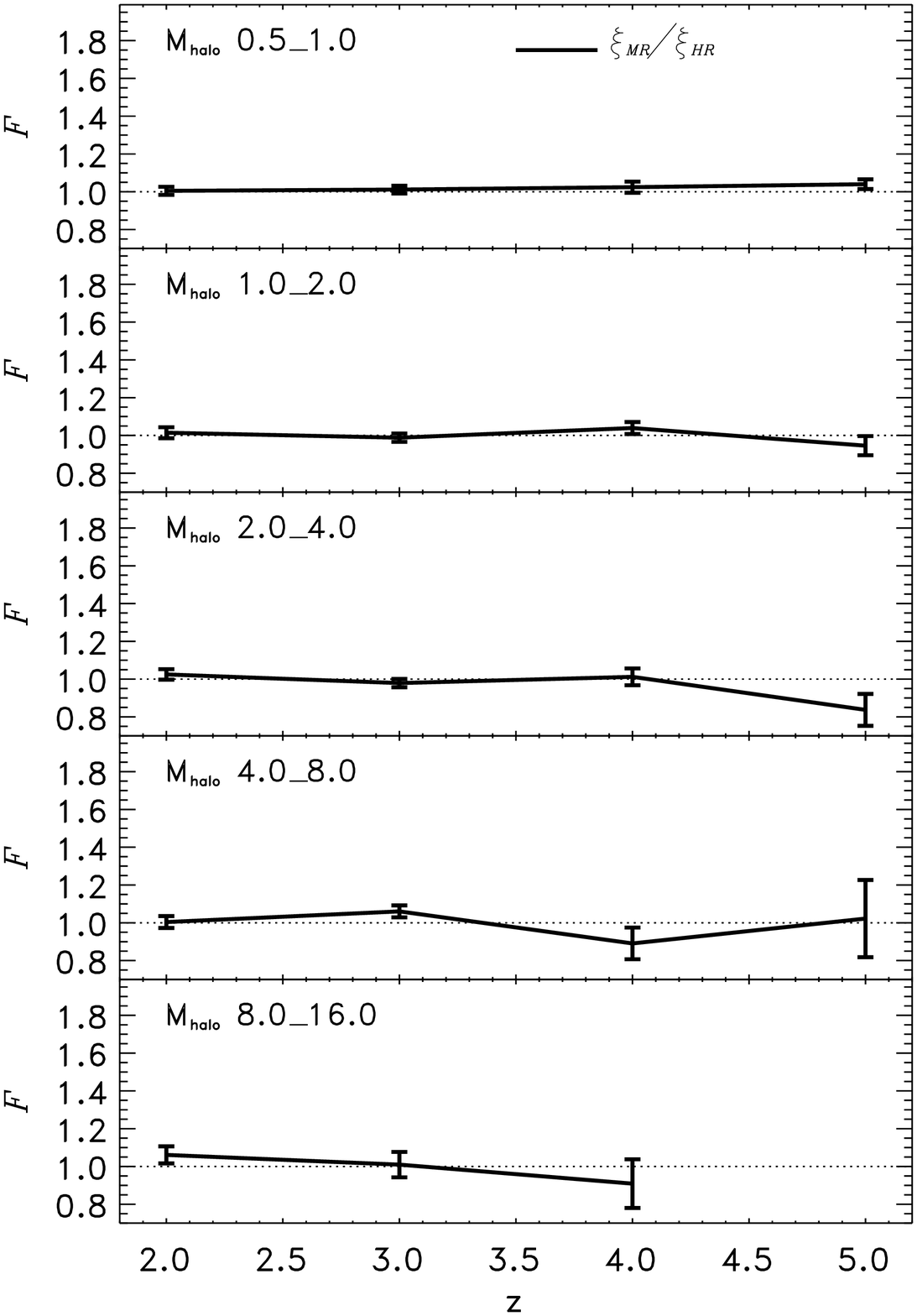}
        \caption{Excess bias for DM haloes in separate mass bins, as
          indicated in each panel in units of $10^{12} \, h^{-1} \,
          \rm{M}_{\odot}$.}
        \label{fig:F_gerard_haloes}
\end{center}
\end{figure}

We estimate errors for our measurements using the bootstrap method,
generating for each sample $50$ bootstrapped samples of the same size,
drawn at random from the parent sample and allowing for repetitions (the
error estimates converge already when using just a few dozen bootstrap
samples). For each bootstrap sample, we calculate the correlation
functions, the bias and the excess bias. The standard deviation among
these quantities is then taken as error estimate.  Recently,
\citet{norberg08} pointed out that the variance on the two-point
correlation function is overestimated when calculated with bootstrap
techniques.  Keeping this in mind, we deliberately choose the bootstrap
method in order to be conservative in our error estimates.  Another
option would have been to estimate errors by subdividing the whole
Millennium volume into subvolumes (for example eight octants) and then
to calculate the variance of the $\xi(r)$ measured within individual
subvolumes. This method becomes inaccurate at large scales (few tens
of $\rm{Mpc}$) due to the smaller volume probed by each subvolume.


\subsection{The merger bias for DM haloes}
\label{sub:haloes}

In our study of the merger bias for haloes we proceed as follows:
\begin{itemize}
\item We take all FOF haloes with mass in the range $5 \times 10^{11} <
  {M}_{\rm Halo} < 1.6 \times 10^{13} ~ h^{-1} \, \rm{M}_{\odot}$. For
  the redshifts analyzed in this work, this mass range is well above the
  collapsing mass $M_{*}$, defined by: $ \sigma(M_{*}) = 1.69$ (at
  $z=2$, $M_{*} \sim 1.3 \times 10^{10} ~ h^{-1} \rm{M}_{\odot}$). This
  entire sample is used as reference sample for the cross-correlation
  analysis. It is large enough that the error on its auto-correlation
  can be safely neglected with respect to other sources of error in the
  $b$ and \F\ parameters (it is composed of $\sim 3.5 \times 10^4$ haloes at $z=5$
  up to $\sim 5.5 \times 10^5$ at $z=2$).
\item We subdivide this sample into five mass-bins, with constant
  logarithmic spacing $\Delta \log \rm{M_{Halo}} = 0.3 $ (a factor of
  two in mass).  We will refer to these five samples as $H_{i}$ .
\item We then checked which haloes in each of the bins of $H_{i}$ had a
  recent major merger, as described in section \ref{sec:halomergers}.
  The subsamples of recently-merged objects are denoted by 
  $M_{i}$. The fraction of merger remnants is $\sim 10\%$ at $z=2$,
  and increases to $ 15-20\%$ at $z=5$.
  The mass bins are narrow enough that, within each bin, the
  merger remnants and the parent population have effectively the same distribution
  of masses. 
\end{itemize}
For the bootstrap error calculation, we created $50$ samples from each
of the $H_{i}$ halo samples, and from these new samples we extracted the
corresponding catalogues of the recently-merged haloes.

If $\xi_{M_{i}, \, R}$ is the cross correlation between the merger
remnants and the reference sample, and $\xi_{H_{i}, \, R}$ is the cross
correlation between all the haloes in the bin and the reference sample,
the excess bias parameter is given by $\mathit{F} = \xi_{M_{i}, \,
  R}/\xi_{H_{i}, \, R}$. One of our principal aims is to quantify how
much $\mathit{F}$ deviates from unity.

In Figure \ref{fig:correlation_haloes}, we show at two different
redshifts an example of the two-point cross-correlation function, the
bias and the $\mathit{F}$ parameter for haloes with masses in the range
$2.0 < M_{\rm Halo} < 4.0 \, \times 10^{12} \, h^{-1}\, \rm{M}_{\odot}$.
In the top panels, the red and blue curves are the cross-correlation
functions between the reference sample and the merger and parent halo
samples, respectively. The error bars show the $1\,\sigma$ dispersion of
the bootstrap samples. The correlation power drops at small scales as
expected for FOF haloes (by definition, two haloes cannot be closer than twice
their virial
radius, hence the ``$1-$halo'' term, i.e., the contribution
to the correlation function from subhaloes within the virial radius, is
missing).  In the middle panels, we show at each scale the bias of the
merger sample, of the parent sample and of the reference sample (red
bow-ties, blue triangles and black squares, respectively); in the lower
panels the excess bias $\mathit{F} = \xi_{M_{i},\, R}/\xi_{H_{i}, \,
  R}$ is also shown as a function of scale.  The errors on each point on
$b$ (and $\mathit{F}$) are from the $1\,\sigma$ dispersion of the bias
($\mathit{F}$) calculated for each bootstrap sample. Both for $b$ and
$\mathit{F}$ the horizontal lines show the best constant fits to the
points in the range $5 < r < 25 ~h^{-1} \rm{Mpc}$.

The resulting fits for the bias and $\mathit{F}$ as a function of
redshift are shown in Figure \ref{fig:bias_f}. The errors on these fits
 are given by the $1\,\sigma$ dispersion of the fits calculated for each
bootstrap sample. The excess bias $\mathit{F}$ corresponding to each
halo mass bin considered is shown in
Figure~\ref{fig:F_gerard_haloes}. If at a given snapshot there are less
than $10$ mergers, we do not show the result, since the corresponding
cross-correlation function would be too noisy.  That is why for the
higher mass-bins (lower panels) the results are not shown at all
redshifts.

In these results, we do not find any statistically significant  
merger bias, over the full redshift range probed in our analysis. Only
in the most massive bins (lower panels), we see a possible signal at the
$\lesssim 5\%$ level for the smallest redshift.  We stress that switching to the \citet{genel08}
catalogs or changing our mass definition does not alter the results presented in the Figures discussed
above.

\begin{figure*}
\begin{center}
	 \includegraphics[width=0.8\textwidth]{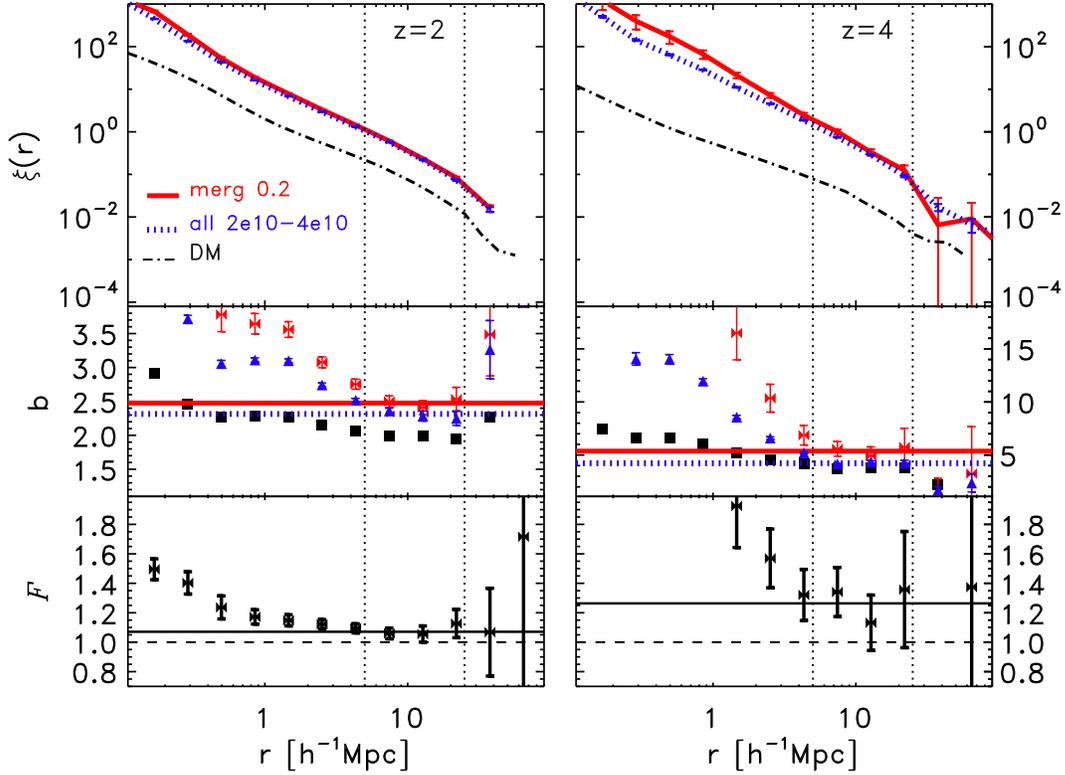}
	\caption{\emph{Upper panels}: Examples of the two-point
          cross-correlation function at $z=2$ and $z=4$ (as labeled)
          between the reference sample and galaxies with stellar
          mass in the range $2.0 < M_{\rm star} < 4.0 \, \times
          10^{10} ~ h^{-1} \rm{M}_{\odot}$ (blue-dotted lines), and
          between the reference sample and the sub-sample of similar mass
          merger remnants (red lines). The auto-correlation of the
          underlying dark matter is shown as dot-dashed line.
          \emph{Middle panels}: Bias as a function of scale for all
          the galaxies in the mass bin (blue triangles), for the
          corresponding merged galaxies (red bow-ties), and for the
          reference sample (black squares).  The horizontal lines
          show  fits to the points, over the range indicated by the
	  vertical dotted lines.  \emph{Lower panels}: Excess
          bias $\mathit{F}$ for the merged galaxies relative to the whole
          galaxy population as a function of scale. The horizontal
          dashed line indicates an absence of excess bias (i.e., 
          $\mathit{F}=1$).  We refer the reader to the text for
          details of the calculation of errors and the fitting
          procedure.}
	\label{fig:correlation_galaxies}
\end{center}
\end{figure*}

\subsection{The merger bias for galaxies}
\label{sub:galaxies}

We investigated the merger bias of galaxies with a procedure similar to
the one adopted above for dark matter haloes.  All the galaxies with
stellar mass in the range $5 \times 10^9 < {M}_{\rm star}< 1.6 \, \times
10^{11} ~h^{-1} \rm{M}_{\odot} $ have been divided into five mass bins,
$G_{i}$, and from each bin we extracted subsamples of
recently-merged galaxies $M_{i}$, obtained as described in section
\ref{sec:galmergers}. We use as reference sample the entire galaxy
population in this range ($5 \times 10^9 < M_{\rm star} < 1.6 \, \times
10^{11}~h^{-1} \rm{M}_{\odot} $, which is composed of $\sim 10^5$
galaxies at $z=5$ up to $\sim 1.4 \times 10^6$ at $z=2$).  This sample
is again large enough that the error on its auto-correlation can be
safely neglected.

In Figure~\ref{fig:correlation_galaxies}, we show an example of the
two-point cross-correlation function for the intermediate mass bin. We
refer to the description of Figure \ref{fig:correlation_haloes} for
details on the derivation of the bias $b$ and the excess bias
$\mathit{F}$.  Unlike for FOF groups, where it only makes sense to
consider the clustering properties on large scales, for the galaxies we
can compute the correlation function down to very small scales $\sim
0.01 ~ h^{-1} \rm{Mpc}$, allowing for a rather accurate description of the 1-halo
term as well, at least at $z<4$.  We find that while \F\ at large scales
is approximately constant, it steadily increases at the smallest scales
probed by our study.
The detection of a steady increase of the excess bias  $\mathit{F}$  with
decreasing scale might be of potential interest. A comparison between
model predictions and the observed small-scale clustering of quasars at
different redshift and luminosity thresholds could, in fact, provide  
insights on the merger-driven nature of quasars, and we postpone a detailed
analysis of this subject to future work.

\begin{figure}
\begin{center}
	\includegraphics[width=0.49\textwidth]{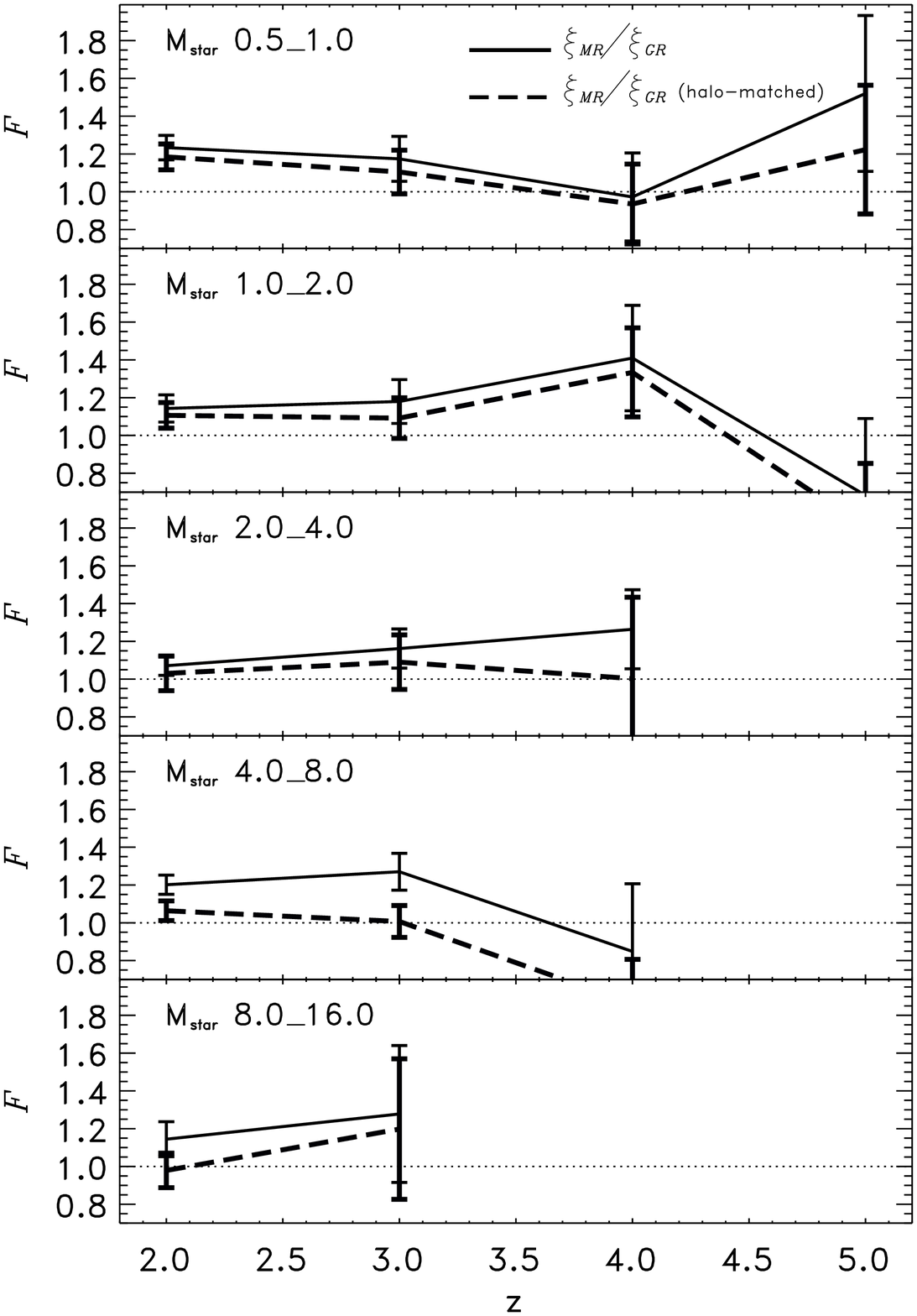}
        \caption{Excess bias between merger remnants and the parent
          galaxy population (solid lines), for different stellar mass bins (the mass
          range is indicated in the upper-left corner of each panel in
          units of $10^{10} h^{-1} \rm{M}_{\odot}$). The dashed lines show the
	  excess bias after  matching the distribution of host
          subhalo masses. No excess bias is
          present if $\mathit{F} = 1$ (thin dotted line). }
        \label{fig:F_basic_halo_gal}
\end{center}
\end{figure}

The excess bias $\mathit{F}$ fitted on scales larger than $5 ~ h^{-1} \rm{Mpc}$  is
plotted as a function of stellar mass and redshift in
Figure~\ref{fig:F_basic_halo_gal} (solid lines).  Excess bias of
size $ 20-30 \%$ ($\mathit{F} \sim 1.2-1.3$) is  clearly present for all mass bins
and at all redshifts, despite the large error bars at the highest
redshifts.  In essence, we find that, at fixed stellar mass, recently
merged galaxies are more strongly clustered on large scales.

\begin{figure}
\begin{center}
	\includegraphics[width=0.47\textwidth]{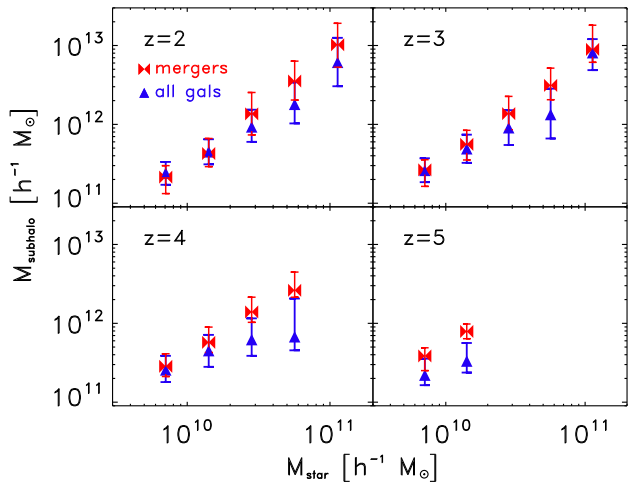}
        \caption{Median host DM subhalo mass corresponding to different
          stellar masses at different redshifts. The red bow-ties
          correspond to haloes hosting recently merged galaxies
          whereas blue triangles refer to the corresponding parent
          population. The error bars represent the $25$ and $75$
          percentiles of the distribution.}
        \label{fig:host_haloes}
\end{center}
\end{figure}

We also examined the mass-distribution of the dark matter subhaloes
hosting the galaxies considered in this analysis. For galaxies that sit
in the main halo of a FOF group, this mass is given by the virial mass
of the group (defined as the mass within the radius that encloses a mean
overdensity of $200$ times the critical density of the simulation),
whereas for galaxies that are located in substructures, the parent
halo mass is defined simply as the number of particles bound to
the substructure (as determined by the {\small SUBFIND} algorithm).  We
found that, for each galaxy bin, the distribution of masses of the host
subhaloes is typically log-normal, and peaks at systematically higher
subhalo masses for recently-merged galaxies.  The median host subhalo
masses for the stellar-mass bins  are
shown  as a function of redshift in Figure~\ref{fig:host_haloes}.

This raises the natural question of whether the excess bias detected for
galaxies could be due simply to this offset in the typical mass of the host subhalo
population. To test this idea, we generated for each galaxy bin a new
parent galaxy population with the same distributions of stellar mass {\it and} host subhalo
mass. The excess bias between this ``corrected'' galaxy population and
the corresponding recently-merged galaxies is shown in
Figure~\ref{fig:F_basic_halo_gal} as a dashed line. 
This exercise significantly decreases the
excess bias signal, and no clear dependence on stellar mass or
redshift remains. Nevertheless, a statistically significant 
excess bias (at the $\sim 10 - 20\%$ level)  still seems to be present,
especially for the lower stellar-mass bins.

In summary, while for FOF dark matter haloes we did not find any
statistically significant  merger bias, for galaxies a
signal is present at a level of $\sim 10 - 20\%$ for the smallest systems.
However, when we restrict ourselves to galaxies at the center of FOF groups ($\sim 75-85
\%$ of galaxies at $z=2$ and $\sim 95-98 \%$ at $z=5$, depending on stellar
mass),
 the excess bias approaches that obtained for dark
matter haloes alone (section \ref{sub:haloes}).  The differing results
obtained for haloes and the galaxy population must therefore be due to
the physics of the merger of galaxies, which goes beyond that of halo merging.
This is still a topic of active research in galaxy formation modelling.

\section{Implications for the clustering of quasars}
\label{sec:qso_clustering}

\begin{figure}
\begin{center}
	\includegraphics[width=0.48\textwidth]{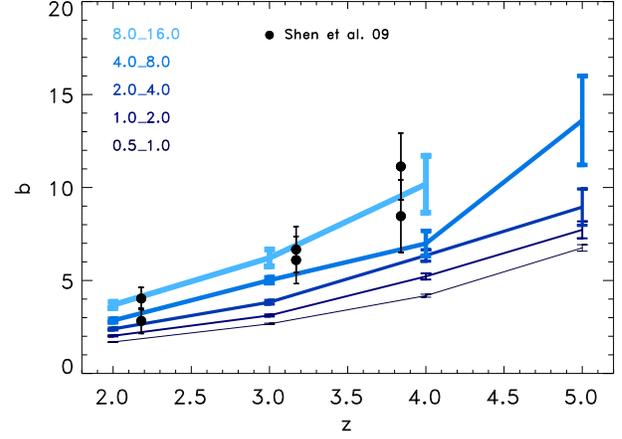}
	\caption{Bias of FOF halo merger remnants as a function of redshift for different mass
          ranges (as indicated on the plots, in units of $10^{12} h^{-1} \rm{M}_{\odot}$). The symbols indicate the
          bias of bright quasars calculated by \citet{shen09},
          inferred from the clustering observations of
          \citet{shen07}.}
	\label{fig:bias_haloes_QSO}
\end{center}
\end{figure}

The large clustering amplitude of quasars observed by \citet{shen07}
at high redshift appears to suggest that these objects live in very
massive haloes.  In Figure~\ref{fig:bias_haloes_QSO}, the bias
associated  with FOF halo merger events for different mass ranges
and at different redshifts, is compared to the observed quasar
bias\footnote{To correct for the different cosmologies used, the large
  scale quasar bias measurements from \citet{shen09} have been
  multiplied by $D_{\rm Shen}(z) ~\sigma_{8,~ 0.78}/(D_{\rm Mill}(z)~
  \sigma_{8,~0.9})$, where $D_{\rm Shen}$ and $D_{\rm Mill}$ are the
  growth factor calculated for the cosmology used by \citet{shen09}
  and the Millennium cosmology, respectively.}, as calculated by
\citet{shen09}. The high observed clustering is compatible
with the clustering associated with the most massive DM haloes, which, at least
up to $z\sim4$, we find to be in better agreement with the analytical predictions of
\citet{jing98}, rather than those of \citet{sheth01}, though still somewhat
higher at the highest redshifts. 

As discussed in Section~\ref{sec:intro}, the high observed clustering
signal forces quasar models to adopt extreme values for some of the
relevant parameters, such as assuming very low scatter in the
$L$-\mh\ relation, high duty cycles, and high radiative efficiencies
\citep{White08,ShankarCrocce}. However, an excess bias applying
specifically to quasar hosts compared with random haloes of the same
mass might reduce  the need for such strong assumptions
\citep{wyithe09}. The results of  the previous
sections for massive haloes and galaxies represent a challenge to this attractive
explanation, at least if the excess bias is to be of merger origin.
Figure~\ref{fig:bias_haloes_QSO} suggests that quasars at $z>2$
 live in haloes $\sim 10^{13} ~h^{-1}
\rm{M}_{\odot}$ which is broadly
consistent with the average host halo mass estimated for lower
redshift quasars \citep[e.g.,][]{croom05}.

If bright quasars have no significant excess bias due to their merger-driven
nature, as our results suggest, then either there is another unknown source
of excess bias, or, more simply, their clustering must trace
the clustering of their host DM haloes and the discrepancy mentioned
above must be explained in some other way. At this point it is therefore
important to carry out a simple consistency check to see if
there are enough massive haloes to host the luminous
quasars actually observed in SDSS at $z\gtrsim 3$.

\begin{figure}
\begin{center}
	\includegraphics[width=0.48\textwidth]{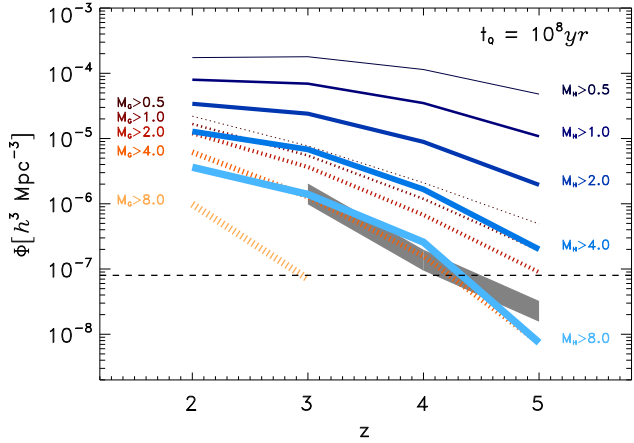}
        \caption{Number density of observed bright quasars from
          \citet{shen07} (gray line), compared with the number of
          major mergers in the Millennium Simulation, obtained by
          multiplying the merger rate by a quasar lifetime
          $t_{q}=10^{8}\rm{yr}$.  The solid blue lines refer to the
          cumulative number density of halo mergers, whereas the red
          dotted lines show the cumulative number of galaxy
          mergers. The minimum mass corresponding to each line is
          shown in the plot in units of $[10^{10} h^{-1}
            \rm{M}_{\odot}]$ for the galaxies and in units of
          $[10^{12} h^{-1} \rm{M}_{\odot}]$ for the haloes.  The
          number densities quoted by \citet{shen07} have been
          multiplied by a factor of two to account for 
          objects missing from optical surveys due to obscuration. }
        \label{fig:number_densities}
\end{center}
\end{figure}

In Figure~\ref{fig:number_densities}, we compare the number density of
observed high-redshift quasars from \citet{shen07} with the number
density of major halo mergers in the Millennium Simulation. Note that
the information extracted from the simulation is a \emph{rate} of
mergers, i.e., the number of merger events within the time interval
between two snapshots (see
Section~\ref{sec:method}). Therefore, when comparing with quasar
number densities we are forced to assume a quasar optical visibility
time $t_{\rm{q}}$, that several independent studies have constrained
to be relatively short and of the order of $t_{\rm{q}}\sim 10^7-10^8$
yr \citep[e.g.,][and references
  therein]{Shankar04,Marconi04,Martini04,YuLu04,Bird08}.  In
Figure~\ref{fig:number_densities} we choose to multiply the rates by a
quasar visibility time of $10^8 \rm{yr}$, which is, at those redshifts, the approximate time
between two snapshots of the Millennium Simulation. The resulting cumulative number
densities are plotted in Figure~\ref{fig:number_densities} and are compared with
the \citet{shen07} quasar number densities. The latter, taken from
Table $5$ in \citet{shen07} and converted to our cosmology, are shown
with a grey band in Figure~\ref{fig:number_densities}, which takes
into account a factor of two uncertainty due to possible sources not
visible in optical surveys due to obscuration. 

From this plot we conclude find that if $t_{\rm{q}}\gtrsim 10^8$ yr, there are potentially enough mergers of
massive haloes in the Millennium Simulation to match the number
density and the large-scale clustering properties of $z>3$ quasars.
A merger model would require a fraction 20-25\% of haloes with mass
$\gtrsim 8\times 10^{12}~h^{-1} \rm{M}_{\odot}$  to be  active at $3<z<4$, in nice
agreement with the analytical models of \citet{ShankarCrocce}, who
find a duty cycle of $0.2-0.4$ within the same redshift range. We stress here
that our mapping between haloes and quasars neglects any  scatter
between halo mass and quasar luminosity, which could spoil the agreement as
discussed by  
\citet{White08}. Significant scatter
in the $L_{\rm QSO}-M_{\rm halo}$ relation would decrease the bias of quasars,
since many would be
hosted by  lower mass (and hence less clustered) haloes. 

The red-colored, dotted lines in Figure~\ref{fig:number_densities}
mark instead the cumulative number densities of galaxy major mergers
 above different final masses, as labeled. The galaxy model predicts that,
on average, the most massive galaxies that recently merged ($M_G
\gtrsim 8\times 10^{10} ~h^{-1} \rm{M}_{\odot}$) reside in the most
massive haloes of mass $\gtrsim 8\times 10^{12} ~h^{-1}
\rm{M}_{\odot}$, with a stellar-to-halo mass ratio consistent with the
one empirically inferred from the cumulative number matching between
the stellar and halo mass functions
\citep[e.g.,][]{Vale04,Shankar06,ConroyWechsler,moster09}.  However,
we find that the number of major mergers for such massive galaxies is
below the number of major mergers of the corresponding hosts, as
clearly seen in Figure~\ref{fig:number_densities} when comparing
dotted to solid lines.  

This apparent discrepancy is explained, first
of all, by the fact that when two haloes merge, their host galaxies
will merge at some later time only if the new satellite halo loses enough mass to fall
below the resolution limit of the simulation. Moreover, in the current
treatment of galaxy mergers, when a galaxy becomes a satellite, it loses its hot gas component; cooling is then
inhibited and the stellar component  grows only moderately from the
cold gas previously available.  Therefore, although a given FOF halo
merger may be counted as a major merger, by the time the corresponding galaxy
merger occurs it may  fall below our chosen
threshold for a major merger. In any case, despite the fact that
the number of major mergers is lower for galaxies than for haloes, the
number of mergers of galaxies more massive than ($M_G \gtrsim 4\times
10^{10} ~h^{-1} \rm{M}_{\odot}$) is still large enough to explain the
observed number densities of bright quasars.

Taken at face value, the galaxy model would then predict that the SDSS
luminous quasars detected at $z>2$ should be hosted by galaxies as
massive as $\gtrsim 4\times 10^{10}~h^{-1} \rm{M}_{\odot}$. Given that
virial relations point to BHs more massive than $\gtrsim 3\times
10^{8} \rm{M}_{\odot}$, this would suggest an increase,
by a factor of $\gtrsim 3$, of the BH-to-stellar mass ratio
with respect to local values \citep{HaringRix}. 
In addition, we find that the clustering of
galaxies with stellar mass $M_{G}\geq 4 \times 10^{10} ~h^{-1} \rm{M}_{\odot}$
is too weak to match the observed quasar clustering.

\begin{figure}
\begin{center}
	 \includegraphics[width=0.47\textwidth]{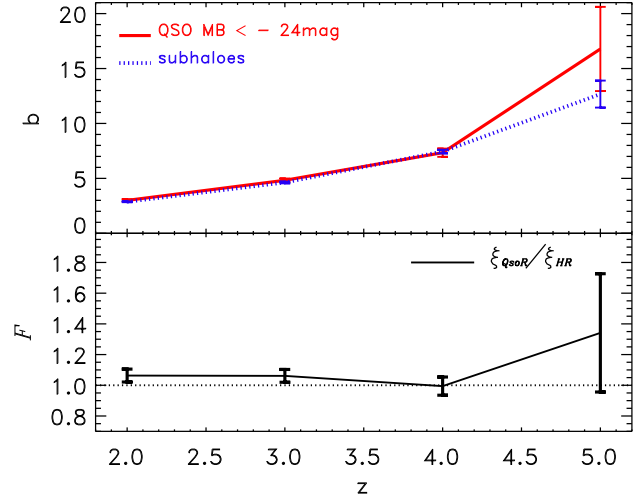}
	\caption{{\em Upper panel:} Bias of simulated bright quasars (B-band
          magnitude $< -24$ mag),
          compared with the bias of randomly selected subhaloes with the
          same mass distribution as the ones hosting the quasars. {\em
            Lower panel:} Excess clustering between the two
          populations.}
	\label{fig:bias_QSO}
\end{center}
\end{figure}

To better address the connection with the semianalytical galaxy
models, we compare our results with the outputs of the detailed model
for the coevolution of quasars and galaxies presented in \citet{marulli08}
and \citet{bonoli09}.  Figure~\ref{fig:bias_QSO} shows the bias of
luminous optical quasars derived using the model of \citet{marulli08},
built on the assumption that quasar activity is triggered during
galaxy mergers. In \citet{bonoli09} we showed that such a model 
predicts  well the clustering properties of observed optical quasars at a
variety of redshifts and luminosities, independent of the
specific  light curve characterizing the active phase of a
BH.  In the upper panel of Figure \ref{fig:bias_QSO}, the bias of
bright quasars in the model is compared with the bias of randomly selected
dark-matter subhaloes with the same mass distribution as the ones
hosting the quasars. The ratio between the two-point correlation functions
of the two samples is shown in the lower panel: the excess bias is at
most $\sim 5\%$, except at $z=5$, where  the small number of
simulated quasars results in a statistically unreliable result.

It is clear that if bright quasars were hosted by DM subhaloes less
massive than  inferred from the clustering analysis, the
BH-to-stellar mass ratio would be even higher (see also the discussion
in \citet[e.g.,][]{ShankarCrocce}). To address, in an independent way,
 the evolution of the average
expected relation between the BH and its host, we compute the expected
baryonic mass  locked in BHs following the method outlined
by previous authors
\citep[e.g.,][]{Ferrarese02,Ciras05,Shankar06,ShankarMathur,ShankarCrocce}:
first, we map haloes to their appropriate virial velocities $V_{\rm
  vir}$ at a given redshift $z$ applying the virial theorem
\citep[e.g.,][]{BarkanaLoeb}. We then link $V_{\rm vir}$ to the
velocity dispersion $\sigma$ as calibrated in the local universe by,
e.g., \citet{Ferrarese02}, and finally we compute the associated BH mass
via the local $M_{\rm BH}-\sigma$ relation \citep[e.g.,][]{Tundo07}. If we
assume that these BHs are accreting at an Eddington ratio $\lambda\gtrsim 0.5$, 
 we find that, at $z=4$, all haloes above $\sim 5\times
10^{12} h^{-1} \rm{M}_{\odot} $ can indeed host a BH luminous enough
to be recorded in the high-$z$ quasar sample of \citet{shen09}.  This simple
exercise proves that if  quasars are associated to normally biased haloes, the
ratio between BH mass and halo mass could be similar to that observed locally.

\section{SUMMARY AND CONCLUSIONS}
\label{sec:summary}

In the present work we exploited the large halo and galaxy samples
extracted from the Millennium Simulation to test the idea that  ``merger
bias'',  a tendency of recently merged systems to be more strongly clustered on large
scales than typical systems of similar mass, could help reconcile the apparent
discrepancy between the observed abundance and  clustering of high redshift
quasars with those predicted for massive dark haloes. Previous studies have, in fact, shown
that the quasar number density and clustering can be simultaneously explained
theoretically either by models characterized by high duty cycles and negligible
scatter in the $L_{\rm QSO}-M_{\rm halo}$ relation
\citep{White08,ShankarCrocce}, or by models with non-zero scatter {\it and} an
excess bias for the haloes hosting quasars \citep{wyithe09}.
 
We quantify the importance of merger bias at different redshifts and for
different halo mass ranges. Defining as major mergers those events in
which two friend-of-friend haloes of comparable mass merge into a single  
system  between two simulation outputs, we
found that recently merged haloes with masses in the range $5 \times 10^{11}$ to $1.6
\times 10^{13} ~ h^{-1} \rm{M}_{\odot}$  show no significant
  excess clustering when compared to other haloes of 
similar mass.

To connect with physically motivated models of galaxy formation, we
also looked for a possible merger bias among samples of galaxies
selected from semi-analytical models built on  the
Millennium Simulation \citep{delucia07}.  We considered galaxies with
stellar mass in the range $5 \times 10^{9} - 1.6 \times 10^{11} h^{-1}
\rm{M}_{\odot}$ and found that merger remnants are typically $10-30
\%$ more clustered than other galaxies of the same mass. 
 The merger remnants are 
hosted by systematically more massive subhaloes than other galaxies, 
explaining a substantial part of this signal. However, even after correcting for
this, we still observe excess
bias at the level of $\sim 5 \%$ for the most massive galaxy merger remnants, and of
$\sim 20 \%$ for our low-mass objects, which are insufficiently clustered
to match the high bias of observed quasars.  If we further restrict the
analysis to central galaxies (i.e.~galaxies at the center of a
friend-of-friend group), for which a clear definition of halo mass is
available, the excess clustering is once more diminished , approaching the null
result  obtained for haloes alone.  

The clear result obtained for haloes and massive galaxies indicates that
merger bias is unlikely to be a  viable solution to the apparent puzzle of the
high clustering of high redshift quasars. On the other hand, we have also found
that recently merged massive haloes with $M_{\rm
halo} \sim 10^{13}~ h^{-1} \rm{M}_{\odot} $ could be both
numerous enough  and clustered enough  to match the observed quasar
number density and large-scale bias, if we assume a  quasar visibility
time $t_{q} \sim 1 \times 10^{8} ~\rm{yr}$ and if we assume negligible scatter in the
relation between
halo mass and quasar luminosity.

In conclusion, if major mergers are responsible for triggering quasar activity, 
the lack of significant merger bias requires models to be
characterized by high duty-cycles and 
negligible scatter in the relation between quasar luminosity and halo mass.    




\section*{Acknowledgments}
We thank Raul Angulo, Mike Boylan-Kolchin, Andrea Merloni, Eyal Neistein and Yue
Shen for
interesting discussions and suggestions. A special thanks goes to Gerard Lemson and Shy
Genel for providing the halo merger trees. SB  acknowledges the PhD fellowship
of the International Max Planck Research School and FS acknowledges support from
the Alexander Von Humboldt Foundation.

\bibliographystyle{mn2e}
\bibliography{merger_bias}

\label{lastpage}

\end{document}